\documentclass{elsart}
\usepackage{natbib}
\usepackage{graphics}
\usepackage{graphicx}
\usepackage{epsfig}
\def\etal{et al.}
\def\ca{{\rm Ca\,II H\,\&\,K }}
\def\ie{i.\,e.\,, }
\begin{document}
\runauthor{Barway, Pandey and Parihar}
\begin{frontmatter}
\title{BVR photometry of a newly identified RS~CVn binary star HD 61396}
\author[Raipur]{Sudhanshu Barway\thanksref{X1}\thanksref{X2}}
\author[Raipur]{S. K. Pandey\thanksref{X2}\thanksref{X3}}
\author[Bang]{Padmakar Singh Parihar\thanksref{X4}}

\thanks[X1]{Corresponding author}
\thanks[X2]{E-mail: ircrsu@sancharnet.in}
\thanks[X3]{Visiting Associate of IUCAA}
\thanks[X4]{E-mail: psp@crest.ernet.in}

\address[Raipur]{School of Studies in Physics, Pt. Ravishankar Shukla University, 
Raipur-India}
\address[Bang]{Indian Institute of Astrophysics, Bangalore-India}

 \begin{abstract}
BVR photometry of a recently identified RS~CVn binary star HD~61396, 
carried out during 2001, is presented. The new photometry reveal 
significant evolution  in the shape and amplitude of light curve 
when compared with those reported earlier by Padmakar \etal (2000). 
The traditional two-starspot model has been used to obtain the spot 
parameters from the observed light curve. Changes in the spot area 
and their location on the stellar surface are discernible from the 
extracted parameters from the new photometry.
\end{abstract}
\end{frontmatter}

\section{Introduction}
HD~61396 (SAO~14296,FG~Cam) is identified as  the 
likely optical counterpart  to the X-ray source 1ES~0738+612 in the 
Einstein IPC Slew Survey by  Elvis \etal (1992).  On  the basis of this 
identification,  supported by the detection of strong  \ca emission by 
Schachter \etal  (1996), and its association with a 29 mJy  source of 
radio emission (BWE 0738+6117) in Becker,  White \& Edwards  (1991) 
Catalog  of 6-cm  sources, Schachter \etal (1996) stated it to be an RS~CVn 
system. Published  measurements   by  the   $Hipparcos$  Satellite
(Perryman \etal   1997) have provided the  following basic parameters
for HD~61396: V$=8.024$, B-V$=1.23\pm0.015$, V-I$=1.19\pm0.01$, 
V$_{max}=7.97$, V$_{min}= 8.11$,  variability magnitude 
($\Delta$V)$=0.112\pm0.021$,and parallax $=3.89\pm0.93$  mili-arcseconds 
(which implies a distance of $257^{+81}_{-50}$pc,  and an absolute visual 
magnitude  of $\sim 1$ indicating  that HD~61396  is  a  luminosity class  
III  star, \ie  a giant) with photometric period of 35.31 days. Furthermore,  
in  the 74th  Special Name-List  of Variable stars (Kazarovets \etal  1999), 
the variability type of FG~Cam was refined to  a semi-regular giant or 
super-giant variable of F-K spectral type, with a period in  the range of 
30 to 1100 days. Detailed optical photometric and spectroscopic, X-ray and 
radio continuum study of HD~61396 by Padmakar \etal (2000) suggest that 
HD~61396 is a long-period RS~CVn binary system with photometric period 
31.95 days showing variation of 0.18 magnitude in the Johnson broad band V. 

If the suggestion of Schachter \etal (1996) and Padmakar \etal (2000) that 
HD~61396 is an RS~CVn type active  binary system  is correct, then  it 
would be  expected to show significant changes in   photometric  light 
variation as 
well as in amplitude  due  to   dark  spots   distributed inhomogeneously 
over the stellar  surface of the primary star, similar to behavior noted in 
other confirmed RS~CVn  stars, such as  $\sigma$~Gem  and $\lambda$~And. 
Keeping this objective in mind we further carried out the 
photometric observation of this star during the year 2001 using the 
40-cm Schmidt-Cassegrain Meade LX-200 telescope situated  in the campus 
of Inter University Centre for Astronomy and Astrophysics (IUCAA) in Pune,
 India and results are presented here.

\section{Photometric Observation}
\begin{figure}
\begin{center}
\includegraphics[]{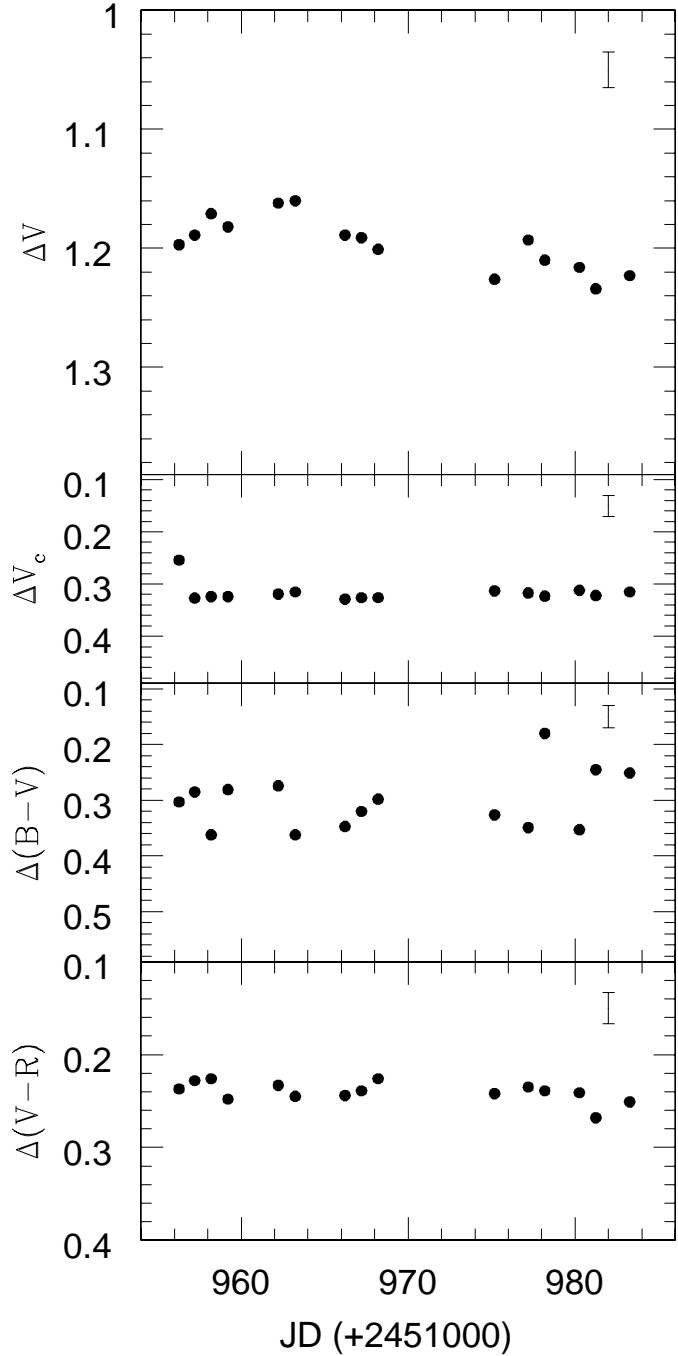}
\caption{V band light curve and (B-V) and (V-R) colors of HD~61396 as observed 
during 2001 using differential photometry plotted against Julian day. V$_c$ is 
for the check star observed on the same night. Typical error bars are shown in 
the upper right corner of each light curve.}
\label{j61_n}
\end{center}

\end{figure}

 The  BVR photoelectric  
photometric  observations of  HD~61396 were carried  out during  the observing 
run during the period February - March, 2001. Because  of rather unfavorable
 sky conditions  we could  observe this star for  a  total of  15 nights only.
 The 40-cm Schmidt-Cassegrain  telescope equipped with SSP-3A 
photoelectric photometer and Johnson standard broad-band BVR filters were used
 for observation. The detector used in SSP-3A photometer was a silicon PN-photodiode
which is not cooled. The response of the B,V and R filters with the detector closely 
matches the Johnson standard B,V and R response function. In order to  obtain 
accurate differential photometry, we used two nearby stars HD~59033 
(K5, V = 6.68, B-V = 0.989) as comparison star and HD~60293
(A0, V = 7.00, B-V = 0.064) as check star. The basic parameters of variable star are 
given in Table 1. The observation were corrected for atmospheric extinction and 
transformed into BVR standard system. The mean  of four  to five independent 
differential magnitudes  measured per night in the V  bands,  and  associated  
colors  (B-V),  (V-R)  are displayed  in figure \ref{j61_n}  as  a function  
of Julian  Day. The uncertainties in $\Delta$V, 
$\Delta$(B-V) and $\Delta$(V-R) are 0.015, 0.02 and 0.017 magnitude respectively. 
We obtained the phase diagram using the photometric ephemeris  
\emph{HJD = 2451209.70 + 31.95$^d$$\times$E} as given in Padmakar \etal  (2000). 

\begin{table*}
\caption{Basic parameters for HD~61396}
%\label{defparagcl} 
\begin{center}
\begin{tabular}{l l l l l l l l l l}
\hline 
Star & V & B-V & Sp. Type & $T_{phot}$ & $\Delta$T & \it i & $\mu_V$ & $\mu_R$  \\
HD~61396 & 8.02 & 1.23 & K2 III & 4520 K & 836 K & 60 & 0.799 & 0.655  \\
\hline 
\end{tabular}
\end{center}
\end{table*}

\section{Results and Discussion}
\subsection{Photometric variability}
\begin{figure*}
\begin{center}
\includegraphics[]{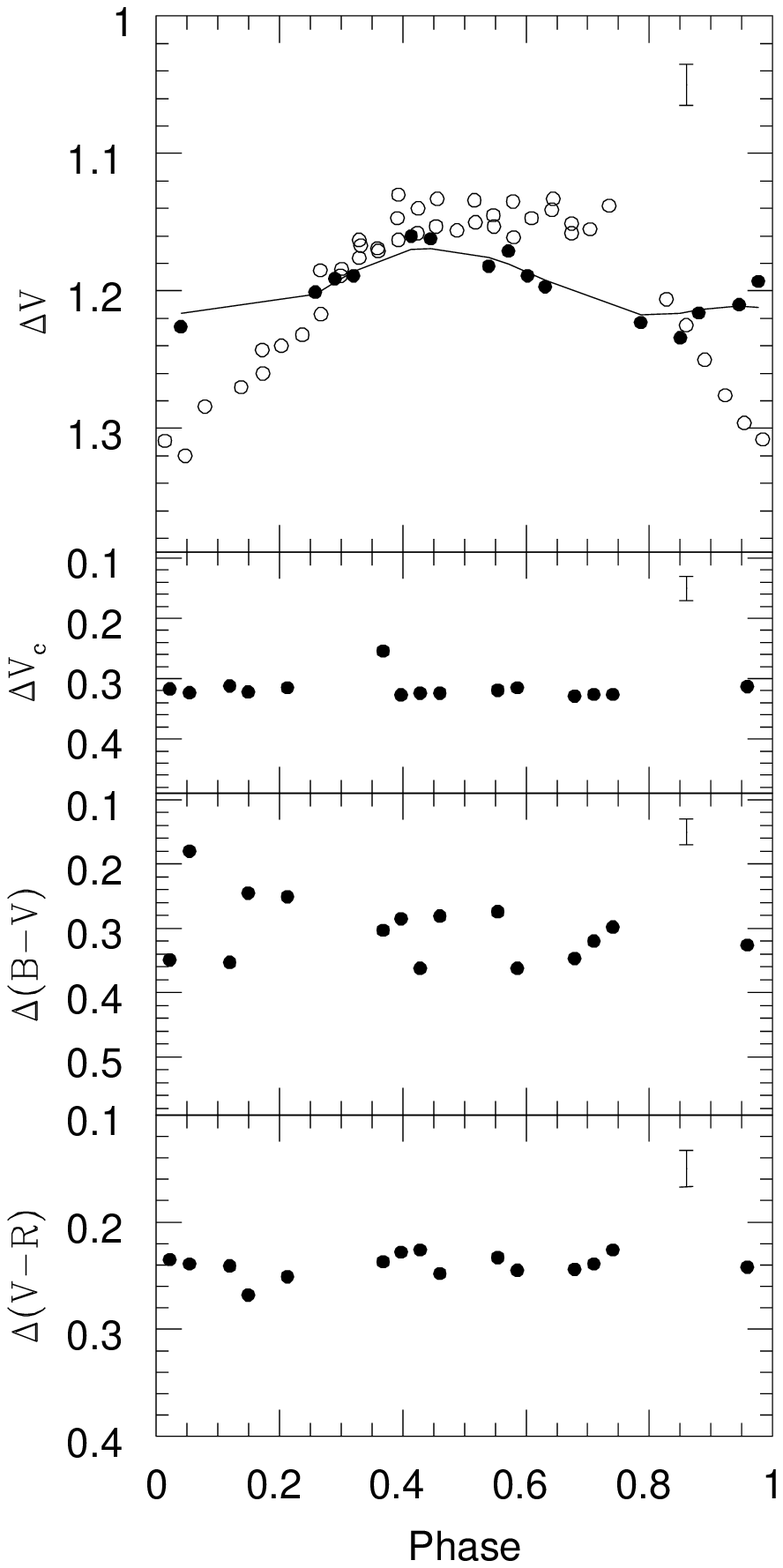}
\caption{As in figure \ref{j61_n} observed data points have been folded in phase 
diagram using the photometric ephemeris \emph{HJD = 2451209.70 + 31.95$^d$$\times$E}. 
The curve represent the best-fit two-spot model for V band data. Open circles denote 
the light curve from Padmakar \etal (2000). Typical error bars are shown in the upper
right corner of each light curve.}
\label{p61_n}
\end{center}
\end{figure*}

Although we have only 15 nights of observations for this star but it is evident 
from the phase diagram (figure \ref{p61_n}) that the observations cover 
the entire phase. In the same figure we also plotted the light curve of HD~61396 
obtained by Padmakar \etal (2000) for comparison. From figure \ref{p61_n}
one can notice a significant light variation as well as change in amplitude 
of light curve from the previous reported  photometry. Present observations 
do not reveal any significant variation in (B-V) and (V-R)color indices. The 
differential amplitude $\Delta V$ turns out to be  $\sim$0.07 magnitude while 
Padmakar \etal (2000) reported a value of $\Delta V$ $\sim$0.18 magnitude.
This indeed further confirms that the star belongs to  class of  RS~CVn 
type variables and the light variation  is due to the presence of starspots 
on the stellar surface.

\subsection{Starspot modeling}
We have modeled the light curve of HD~61396 in  the V band in the frame 
work of starspot. The details of 
modeling technique are  given in Padmakar \& Pandey (1999). To extract
 geometrical  parameter of spots we adopted the analytical formulation given by 
Dorren (1987). We have taken two circular spots on the stellar 
surface and another spot  permanently located on poles for modeling. To model V band 
light curve we have taken wavelength-dependent limb-darking coefficients 
($\mu$), inclination ({\it i}), photospheric temperature (T) and difference 
between the photospheric and spot temperature ($\Delta T$) as constant 
(given in Table 1) and the longitude($\lambda$), latitude($\beta$), 
radius($\gamma$) as free parameters. The final best fit parameters along with 
their uncertainties are listed in Table 2.

Polar spots or spots uniformly distributed over the stellar surface do not 
produce any rotational signature in the light output, and therefore, are 
undetectable by the photometric modeling of the light curve. Our
spot modeling technique for this system indicates that two nearly equal sized spots
 having radii of 10.26$^\circ$ and 11.80$^\circ$ situated at opposite hemispheres
 were responsible for the observed light variation. The spots were separated 
from each other by $\sim$ 230$^\circ$ in longitude, and cover $\sim$ 2 percent of 
the stellar surface. A comparison of our results with those of Padmakar \etal 
(2000) shows that there is  a substantial variation in the location of the 
minimum and the shape of the light curve. This behavior in turn reflects that 
changes in the amplitude are mainly due to the redistribution of spots on the 
stellar surface rather than overall changes in the level of spottedness.
  
\section{Conclusion}
We have presented  BVR photometric observations for a  newly identified
RS~CVn type star HD~61396. The light curve show significant variation in the
shape as well as amplitude. The differential amplitude in V is $\sim$0.07 mag
which is smaller than the previously reported value. The spot parameters 
also indicate changes in their location as well as in size, supporting the
starspot hypothesis. A comparison from previous observations reported by Padmakar
\etal (2000) shows that there is a significant variation in location of the 
minimum and the shape of the light curve and it is quite conceivable that 
the observed light variation is due to a change in the position of the major 
starspot.

\begin{table*}
\caption{Starapot parameters for HD~61396}
%\label{defparagcl} 
%\begin{center}
\begin{tabular}{c c c c c c c c c}
\hline 
\multicolumn{3}{c}{Spot 1} & \multicolumn{3}{c}{Spot 2} & Polar Spot & Total Spot& $\chi^2$ \\
 $\lambda$ & $\beta$ & $\gamma$ & $\lambda$ & $\beta$ & $\gamma$ & Radius & Area (\%) &  \\
285.07 & 35.41 & 10.26 & 53.46 & -2.37 & 11.80 &  38.59 & 1.86 & 1.43 \\
$\pm$ 3.27 & $\pm$1.08 & $\pm$0.40 & $\pm$ 0.94 & $\pm$1.42 & $\pm$ 0.54 & $\pm$ 0.31 & $\pm$ 0.11 & \\
\hline 
\end{tabular}
%\end{center}
\end{table*}

\bigskip
\noindent
{\bf Acknowledgments} \\
We are thankful to IUCAA for providing their observing, library and computing
facilities. We also express our sincere thanks to Professor A. K. Kembhavi for
 his kind co-operation and suggestions during the course of the observations.
We thank the anonymous referee for improving this manuscript with his comments.
SB and SKP would like to express their sincere thanks to CSIR, New Delhi, for
 financial support through a project grant No. 03(0985)/03/EMR-II.

\end{document}